\documentclass[]{spie}  

\usepackage{amsmath,amsfonts,amssymb}
\usepackage{graphicx}
\usepackage[colorlinks=true, allcolors=blue]{hyperref}
\usepackage{xspace}
\usepackage{gensymb}

\newcommand{\ie}{\textit{i.e.}\xspace}
\newcommand{\sref}{Sec.~\ref}
\newcommand{\fref}{Fig.~\ref}

\newcommand{\arcsec}{\ensuremath{^{\prime\prime}}\xspace}

\title{The VLT/ERIS vortex coronagraph: design, pointing control, and on-sky performance }

\author[a]{G. Orban de Xivry}
\author[a]{O. Absil}
\author[b]{R. J. De Rosa}
\author[c,k]{M. J. Bonse}
\author[c]{F. Dannert}
\author[c]{J. Hayoz}
\author[d]{P. Grani}
\author[d]{A. Puglisi}
\author[e]{A. Baruffolo}
\author[e]{B. Salasnich}
\author[f]{R. Davies}
\author[c]{A. M. Glauser}
\author[g]{E. Huby}
\author[h]{M. Kenworthy}
\author[c]{S. P. Quanz}
\author[i]{W. Taylor}
\author[j]{G. Zins}

\affil[a]{Space sciences, Technologies, and Astrophysics Research (STAR) Institute, Université de Liège, allée du Six Août 19c, 4000 Liège, Belgium}
\affil[b]{European Southern Observatory, Alonso de Córdova 3107, Vitacura, Santiago, Chile}
\affil[c]{ETH Zurich, Institute of Particle Physics and Astrophysics, Wolfgang-Pauli-Strasse 27, 8093 Zurich, Switzerland}
\affil[d]{INAF – Osservatorio Astrofisico di Arcetri, Largo E. Fermi 5., 50125 Firenze, Italy}
\affil[e]{INAF – Osservatorio Astronomico di Padova, Vicolo dell’Osservatorio 5, 35122 Padova, Italy}
\affil[f]{Max-Planck-Institut für extraterrestrische Physik, Postfach 1312, 85741 Garching, Germany}
\affil[g]{LESIA,  Observatoire de Paris, Université PSL, CNRS, Sorbonne Université, Université de Paris, 5 Place Janssen, 92195 Meudon, France}
\affil[h]{Leiden Observatory, University of Leiden, PO Box 9513, 2300 RA Leiden, The Netherlands}
\affil[i]{STFC UK ATC, Royal Observatory Edinburgh, Blackford Hill. Edinburgh EH9 3HJ, UK}
\affil[j]{European Southern Observatory, Karl-Schwarzschildstr. 2, 85748 Garching, Germany}
\affil[k]{Max Planck Institute for Intelligent Systems, Max-Planck-Ring 4, 72076 T\"ubingen, Germany}

\authorinfo{Further author information: (Send correspondence to Gilles Orban de Xivry)\\G. Orban de Xivry: E-mail: gorban@uliege.be.}

\begin{document}
\maketitle

\begin{abstract}
The Enhanced Resolution Imager and Spectrograph (ERIS) is the new near-infrared instrument at the VLT-UT4. ERIS replaces and extends the observational capabilities formerly provided by SINFONI and NACO: integral field spectroscopy at 1 - 2.5 $\mu$m, imaging at 1 - 5 $\mu$m with several options for high-contrast imaging, and long-slit spectroscopy.
In particular, a vortex coronagraph is now available for high contrast observations at L and M band. It is implemented using annular groove (or vortex) phase masks (one for each of the L and M bands) in a focal plane, and a Lyot stop in a downstream pupil plane. The vortex coronagraph has a discovery space starting already at $\sim$1$\lambda$/D, and works well in broadbands. However, to reach its optimal performance, it is critical to correct for slow pointing errors onto the vortex phase mask, which mandates a dedicated pointing control strategy.
To do so, a control loop based on the QACITS algorithm has been developed and commissioned for ERIS. Good pointing stability is now regularly achieved with errors between 0.01 and 0.02 $\lambda$/D and a correction rate of 0.2 Hz.
In this contribution, we first review the design of the ERIS vortex coronagraph. We then detail the implementation of the QACITS algorithm describing the entire observing sequence, including the calibration steps, the initial centering, and the stabilization during the observing template.
We then discuss performance based on commissioning data in terms of pointing accuracy
and stability. Finally, we present post-processed contrast curves obtained during commissioning and compare them with NACO vortex data, showing a significant improvement of about 1 mag at all separations.

\end{abstract}


\section{INTRODUCTION}
\label{sec:intro}
The Enhanced Resolution Imager and Spectrograph for the VLT replaces  NACO and SINFONI in a single instrument and with improved diffraction limited and spectroscopic capabilities\cite{Davies+2023}.
The instrument consists of four main building blocks: a calibration unit, the integral field spectrograph SPIFFIER, the imager NIX, and the adaptive optics (AO) wavefront sensor.

NIX provides diffraction-limited imaging from 1 to 5 $\mu m$ and features a comprehensive set of broadband and narrowband filters. Besides classical imaging and long-slit spectroscopy, NIX supports  several high-contrast imaging modes:
\begin{enumerate}
    \item A pupil plane coronagraph based on a grating vector apodised phase plate (gvAPP)\cite{Doelman+21}. This coronagraphic modes generates, for any star in the field, two coronagraphic images with conjugated D-shaped dark holes, each with 10-30\% of the flux, depending on the filter\cite{Davies+2023}. When combined, the gvAPP provides almost 360$\degree$ discovery space from about 2 to 15 $\lambda/D$. Because the separation of the two images scales with wavelength, the gvAPP is  best suited for narrow-band filters.
    \item Three sparse aperture masks provide non-redundant (or partially non-redundant) apertures, yielding a resolution that is a factor of 2 better than the diffraction limit of the telescope. The three masks offer three different trade-offs between $u, v$ coverage, throughput, and non-redundancy.
    \item A focal plane coronagraph (FPC) using vortex phase masks, one for each of the L and M bands, working well in broadband. It provides a clear 360$\degree$ discovery space starting already at $\sim 1\lambda/D$. To perform optimally, however, it requires precise pointing control to keep the star centered on the vortex phase mask.
\end{enumerate}
This contribution focuses on the focal plane (vortex) coronagraphic mode, detailing its design, optimization, and on-sky operation.

One significant enhancement of ERIS over NACO and SINFONI is its improved AO system\cite{Riccardi+22}. The ERIS AO system makes use of the VLT AO Facility including, most notably,  the Adaptive Secondary Mirror (ASM) on  VLT-UT4. The system offers several natural guide star and laser guide star modes, tailored to its various science cases. The wavefront sensor is based on a Shack-Hartman with 40-by-40 subapertures, each with 6\arcsec$\times$6\arcsec field-of-view.
The opto-mechanical design employs only flat or slow optics to relay the telescope beam  minimizing and stabilizing non-common path aberrations (NCPAs). Indeed, during assembly, integration, and verification, the NCPAs were calibrated, showing contribution ranging from 22 to 32 nm rms wavefront error\cite{Riccardi+22}. This low NCPA level is attributed to the use of quasi-plano optics and the telecentric design, enabling NGS acquisition  with stages rather than a field steering mirror.

ERIS, with its extensive capabilities, is the new VLT workhorse instrument for diffraction limited near-infrared observation of individual targets. Key science cases\cite{Davies+2023} include monitoring  stellar proper motions around Sgr A$^*$, studying galaxy evolution at high redshifts  ($z\sim 1-3$) to unveil the physical processes shaping galaxies, and directly imaging self-luminous gas giants and protoplanets as well as probing their atmospheres.
ERIS saw its first light in 2022, with the bulk of the commissioning happening  that year. A wide array of its modes has been offered since ESO Period 111 (for observations beginning in April 2023) and the FPC mode is now offered since ESO Period 113 (for observations beginning in April 2024).

In the context of direct imaging, the vector vortex coronagraph is an attractive solution thanks to its high throughput, small inner working angle down to $\sim 1 \lambda/D$, its 360 degrees clear discovery space, and its achromatic properties.
Over the past decade, mid-infrared vortex coronagraphs have been integrated into several instruments:  VLT/VISIR\cite{Delacroix+12}, VLT/NACO\cite{Mawet+13}, LBT/LMIRCam\cite{Defrere+14}, Keck/NIRC2\cite{Femenia+16}, VLT/NEAR\cite{Maire+20}, and now VLT/ERIS.
However, small inner working angle focal-plane coronagraphs also mean a high sensitivity to pointing drifts, and require specific control to perform optimally.

In this contribution, we review the principle and the design of the ERIS vortex coronagraph. We then describe the pointing control algorithm specifically tailored for the vortex, and  the implementation of the FPC operation into the observing templates. Finally, we present the obtained on-sky results in terms of pointing accuracy, raw contrast, and post-processed contrast.

\section{VLT/ERIS VORTEX CORONAGRAPH}
A vortex coronagraph features a vortex phase mask in its focal plane. The text book effect is to move the light of an on-axis source outside the downstream geometric pupil. A Lyot stop then blocks the diffracted light and
provides theoretically a perfect starlight cancellation for a circular entrance aperture.
However, realistic pupils featuring a central obscuration and spiders lead to  a leakage term not entirely blocked by the Lyot stop, \ie the vortex coronagraph (\ie a vortex phase mask coupled with a downstream Lyot stop) cannot be considered `perfect' any longer.

The ERIS vortex focal plane coronagraph, similarly to all the mid-infrared vortex coronagraphs so far, is implemented using an annular groove phase mask (AGPM\cite{Mawet+05}), followed by its Lyot stop in the pupil plane.
AGPMs are made up of concentric subwavelength gratings etched onto diamond substrates. The subwavelength grating acts locally as an achromatic half-wave plate and, together with the concentric design,  synthesizes a $ 4\pi$ helical (vortex) phase ramp for the two orthogonal polarization states of light separately.
For improved transmission, an anti-reflective grating is etched on the backside of the mask which significantly reduces the reflection from $\sim 17\%$  to $\sim 1.9\%$ (L band)\cite{Delacroix+13}.

ERIS has two AGPM masks, one for each of the L and M bands.
The components were developed and tested by Liège University and manufactured by Uppsala University. The tests were performed on the VODCA test bench\cite{Jolivet+2019} where we measured the component transmissions and rejection ratios. One of the two components was initially under-performing and was subsequently re-etched and tested. The final measurements performed on VODCA, also reported in Jolivet et al. \cite{Jolivet+2019}, are given in Tab. \ref{tab:agpm}.
Both components show excellent rejection performance with ratio of 1773:1 (at 4.6$\mu m$) for the ERIS M band vortex, and 1430:1 (measured on a bandpass of  from 3575 to 4125 nm) for the ERIS L band vortex.

\begin{table}[]
    \caption{Measurements of the AGPM transmission and rejection ratio on the VODCA test bench of the two ERIS AGPM. See also Jolivet et al.\cite{Jolivet+2019} for more details. The L-band measurement are performed on a bandpass from 3.575$\mu m$ to 4.125$\mu m$. }
    \label{tab:agpm}
    \centering
    \begin{tabular}{l cccc}
    \hline
         Name &  Transmission  &Transmission  & Rejection ratio  & Rejection ratio \\
         & (L- broadband) & (4.6$\mu$m) & (L broadband) & (4.6$\mu$m) \\
         \hline
         AGPM-L12r   &  $85\% \pm 0.7$ &
         $68.5\% \pm 2.5$  & $84 \pm 5$ & $1773 \pm 95$\\ 
         AGPM-L15     & $83.5\% \pm 0.5$ & $66\% \pm 1$ & $1430 \pm 82$ & $50 \pm 3$\\ 
         \hline
    \end{tabular}
\end{table}

The downstream Lyot stop blocks the star light in the pupil plane. The stop is typically undersized to block the light diffracted by the secondary obscuration and the secondary support structure, which appears as a monotonically decreasing flux moving away from these structures. The Lyot stop therefore shrinks the perimeter of the entrance pupil in order to mask out the scattered on-axis light. The trade-off consist in reducing the on-axis stellar light leakage but maintaining a sufficient throughput for off-axis sources.

\begin{figure}[p]
    \centering
    \includegraphics[width=0.4\linewidth]{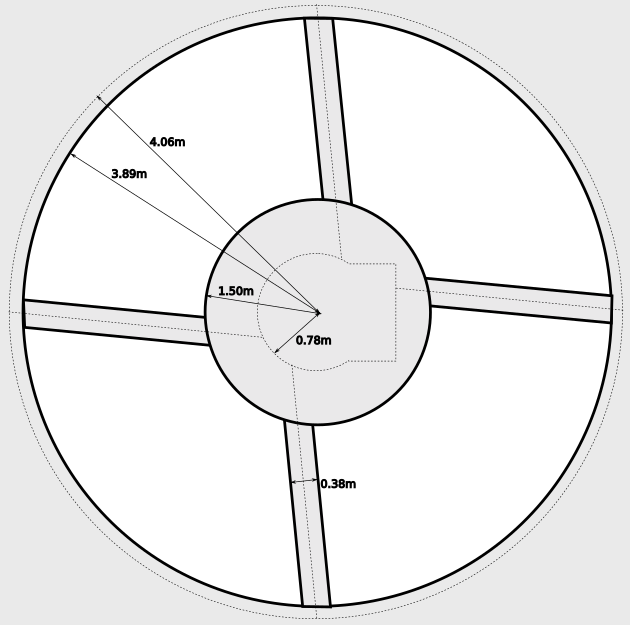}
    \quad
    \quad
    \includegraphics[width=0.4\linewidth]{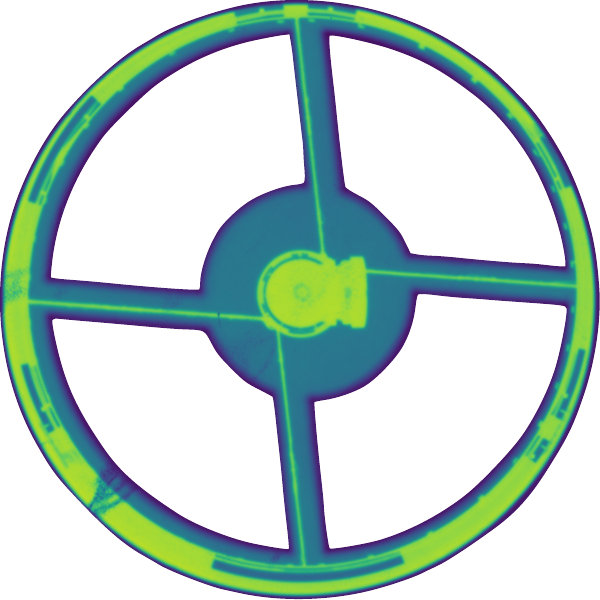}
    \caption{ (Left) Vortex Lyot stop with its dimensions given with respect to  VLT M1. (Right) Composite pupil image obtained with NIX, showing the entrance pupil including the warm support structure of M1 and M2, and the cold Lyot stop with its thick spiders and oversized central obscuration. The four central white areas are the portion of the pupil not blocked by the Lyot stop. }
    \label{fig:pupil}
\end{figure}

The ERIS Lyot mask itself is made from Molybdenum and blackened with Acktar Fractal black coating.
Two stops are installed: the regular Lyot stop, and a Lyot stop coupled with a ND filter allowing the observation of very bright stars.
The final dimensions with respect to the VLT entrance pupil are given in \fref{fig:pupil}a. The outer diameter is 95.7\% of the pupil diameter, the inner diameter is 37\% of the pupil diameter. The throughput of the Lyot stop is 74.77\%.
A composite image of the entrance pupil and the Lyot stop on the VLT-UT4 is shown in \fref{fig:pupil}b. As ERIS does not have a cold pupil stop upstream of the vortex, the warm support structure of the VLT primary mirror and its secondary is visible. With the Lyot stop slid in, we can verify the  alignment of the  stop with respect to the entrance pupil.

\begin{figure}[p]
    \centering
    \includegraphics[width=0.5\linewidth]{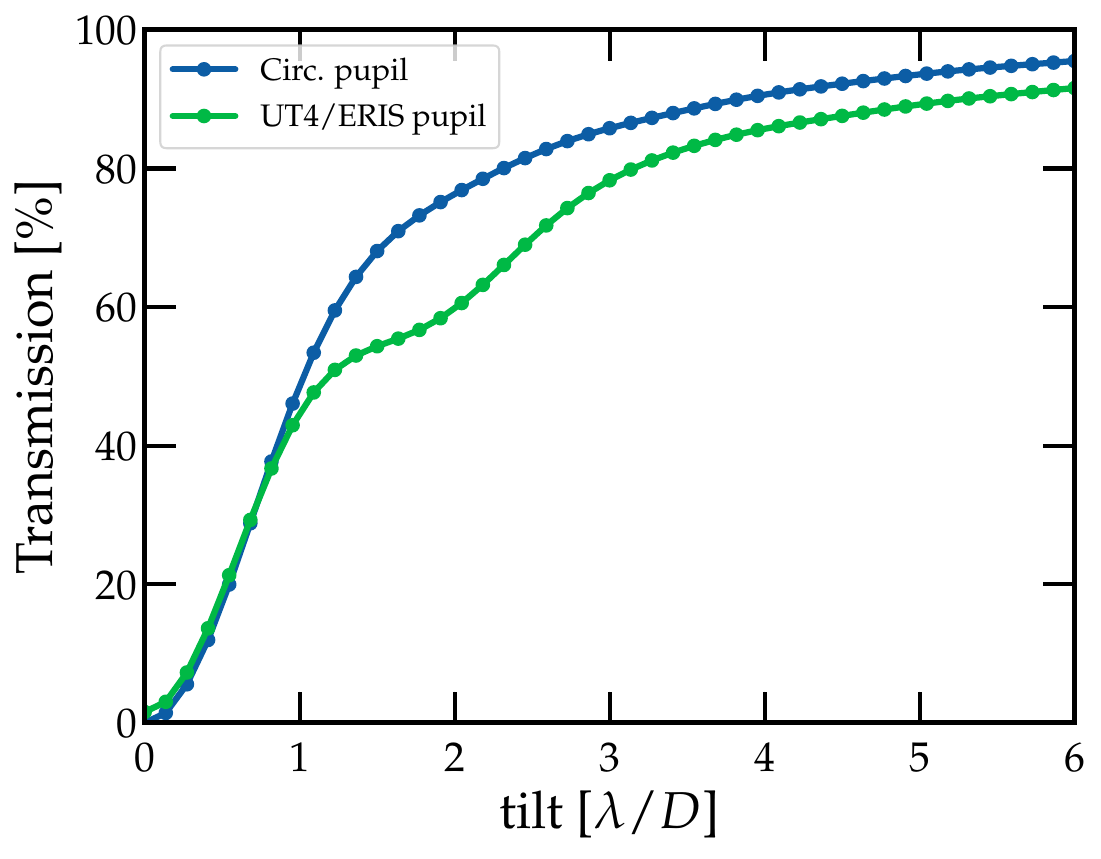}
    \includegraphics[width=1\linewidth]{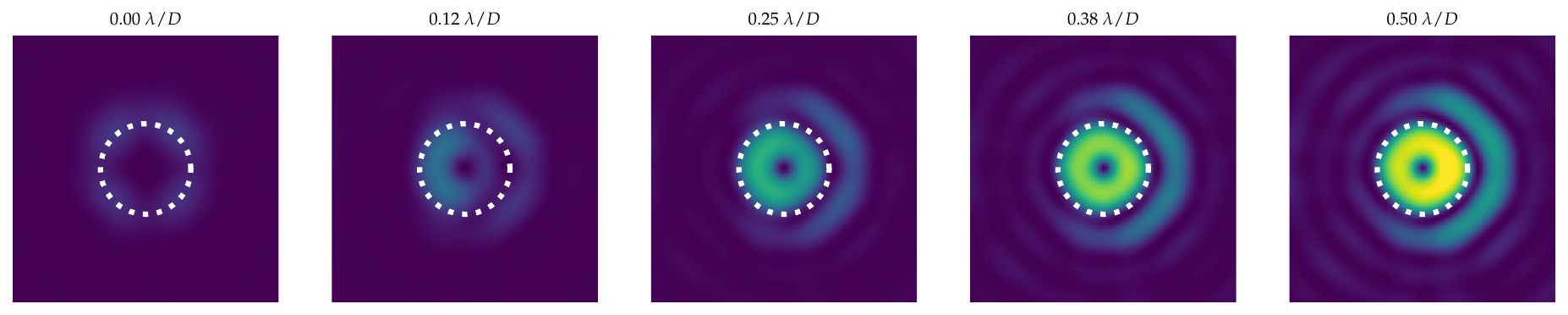}
    \caption{({\it Top}) Theoretical coronagraphic transmission as a function of radial distance to the vortex center, for a circular pupil (blue) and more specifically for ERIS (green). The coronagraphic transmission reaches $\sim 50\%$ at 1 $\lambda/D$ which makes the vortex one of the most efficient in terms of inner working angle, but also more prone to pointing errors ({\it Bottom}) Illustration of the vortex coronagraphic PSFs for decentering from up to 0.5 $\lambda/D$ showing the rapid increase in transmission. The white circle indicates the separation between the inner and outer estimator defined at 1.7 $\lambda/D$ radius.}
    \label{fig:eris-simu}
\end{figure}

The vortex components are mounted in the NIX aperture wheel, and the Lyot stops are mounted in the NIX pupil wheel\cite{Glauser18}.

The combined throughput (vortex phase mask and Lyot stop) of the focal plane coronagraph amounts to $\sim$62\% in L band and $\sim$51\% in M band.
This throughput does not take into account the Strehl ratio reduction due to the Lyot stop, which is undersized compared to the entrance pupil.

\section{VORTEX CENTERING WITH QACITS}
To perform optimally, the star needs to be precisely centered on the vortex phase mask, in which case it suppresses the central PSF significantly, leaving only a ring of residual light.
Indeed, for focal plane coronagraphs, small inner working angles also mean high sensitivity to pointing drifts. This is illustrated in \fref{fig:eris-simu} using Fourier-based simulations of the ERIS vortex coronagraph. The transmission plot shows that at $\sim 1\lambda/D$, almost 50\% of the flux is transmitted. This is also visually illustrated in the image sequence below with tilt going from 0.01 to 0.7 $\lambda/D$.

To keep the star centered on the mask during the observation, we use an automated algorithm, specifically developed for the vortex coronagraph, called  Quadrant Analysis of Coronagraphic Images for Tip-Tilt Sensing (QACITS; Huby et al.\cite{Huby+15}). The QACITS technique is based on a quadrant analysis of the vortex coronagraphic images. Because the relationship between the measured shift and the tilt (or tip) is non-linear and generally not bijective when considering the whole image, QACITS splits down the focal plane images in several areas to build different estimators.
The estimator providing the largest linear range is the so-called `outer estimator'. It only considers the quadrant flux asymmetry in an annulus going from 1.7 to 2.3 $\lambda/D$. The outer estimator is robust to various effects, such as misalignment or low order aberrations, but at the expense of a somewhat lower signal-to-noise ratio (SNR). %

\subsection{On-sky calibration}
To calibrate and validate our QACITS model, we observe a bright target on-sky, manually center it on the vortex phase mask, and then incrementally offset the star from the mask along one axis. We then reproduce this procedure along the other axis.
The offsets are injected by directly  moving the AO stages.

\begin{figure}[t]
    \centering
        \includegraphics[width=0.8\linewidth]{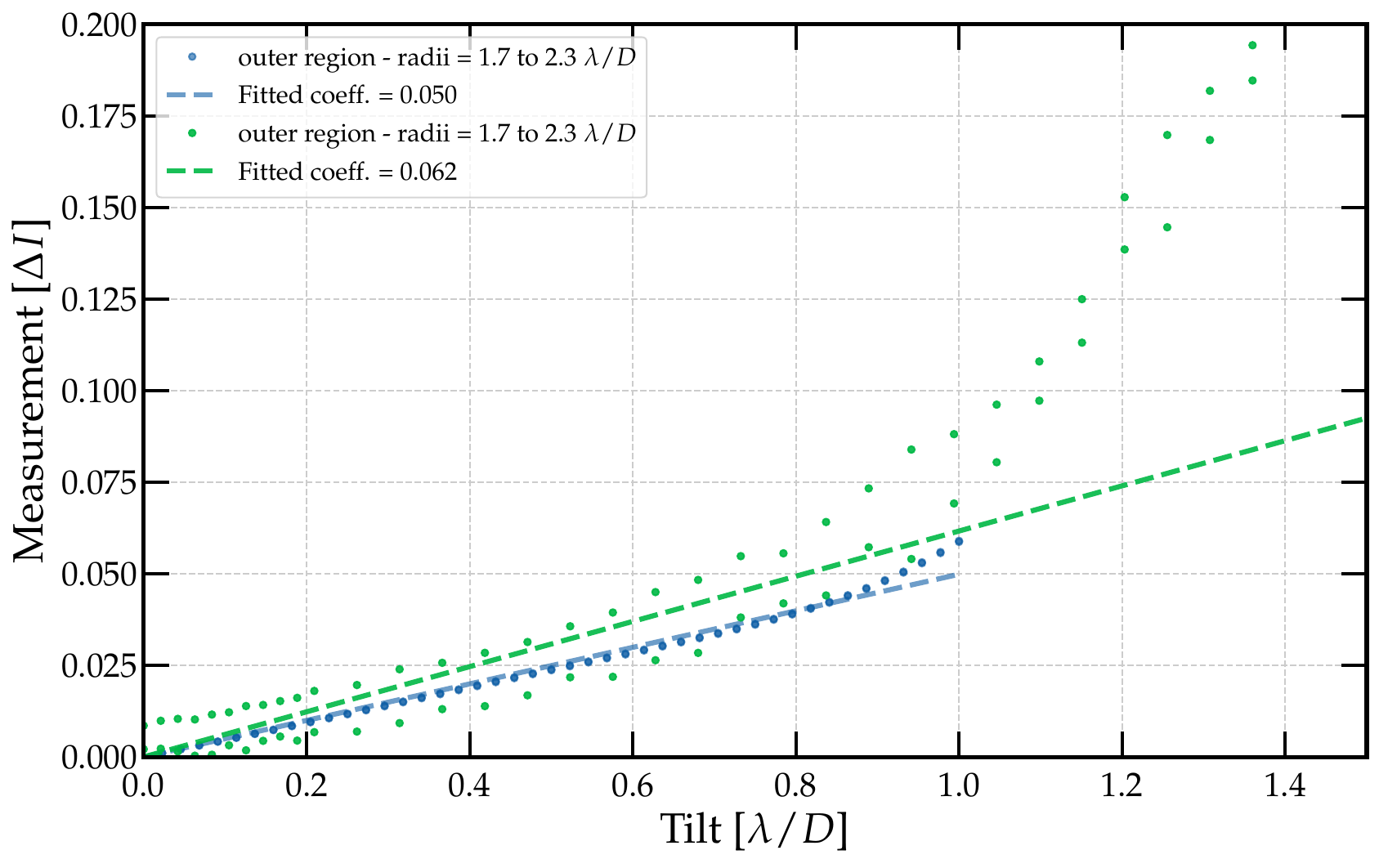}
    \caption{Calibration of the QACITS response using the outer estimator. In green, the $x$ and $y$ measurements obtained on-sky. In blue, the same calibration performed in simulation (up to 1 $\lambda/D$ tilt).}
    \label{fig:calibration}
\end{figure}

Based on those measurements, we can produce the response curves of the QACITS estimator which means defining the slope between measurement and tip-tilt.
The results, shown in \fref{fig:calibration}, indicate a slope of 0.062. Comparing the on-sky measurements to our simulations shows a relatively good match between the response curves, validating our on-sky calibration.
As shown by the curve, the linear range of the outer estimator is approximately $1 \lambda/D$ or $ \sim 0.1''$.

\subsection{Control aspects}
\begin{figure}
    \centering
    \includegraphics[width=0.9\linewidth]{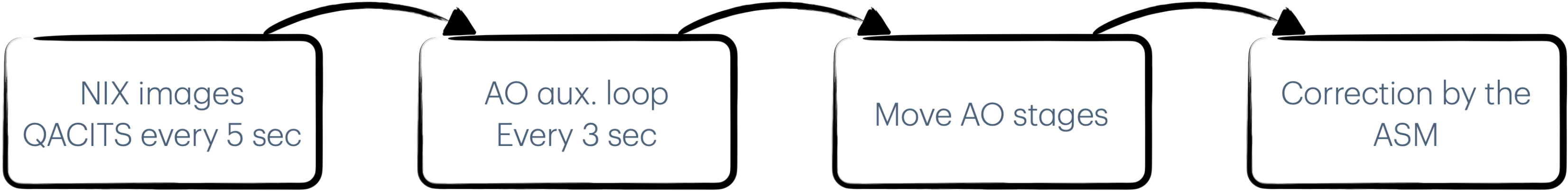}\\
        \vspace{0.5cm}
    \includegraphics[width=0.45\linewidth]{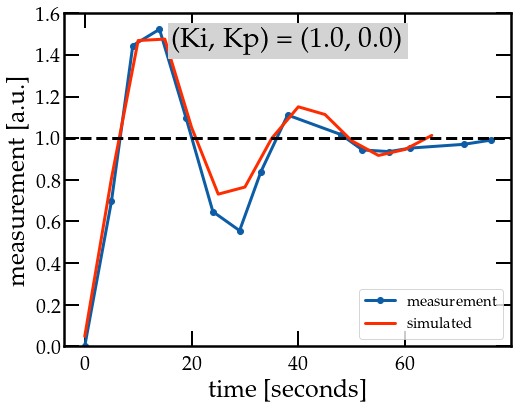}
    \includegraphics[width=0.45\linewidth]{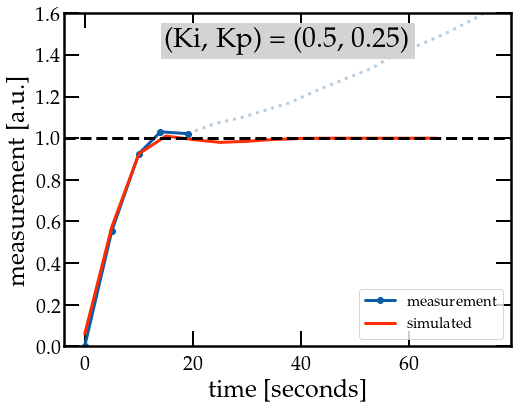}
    \caption{({\it Top}) Control flow diagram. ({\it Bottom}) Step response curves to perturbations of 10 pixels. The left figure shows the response obtained for integrator and proportional gains $(K_i, K_p) = (1, 0)$ and the right picture shows the response after adjustment of the gain to $(K_i, K_p)=(0.5, 0.25)$. In blue, the on-sky measurements and in red, the simulated curves using our toy model. The blue curve of the right plot changes to a dotted light blue at around  20 seconds, indicating that the correction reached a software limit. This is a specific experimental case and is not a concern for regular operation.}
    \label{fig:step_response}
\end{figure}

The control sequence of QACITS is illustrated in \fref{fig:step_response} and works as follows:  QACITS measures the residual drift every 5 seconds based on the available science images, then a total command is computed using a proportional-integral (PI) controller and is sent to an asynchronous auxiliary AO loop with a repetition rate of 3 seconds. The commands sent to the auxiliary AO loop result in a movement of the  AO wavefront sensor stages, which subsequently causes a tip-tilt correction by the adaptive secondary mirror.

The control parameters of the PI controller were optimised on-sky by monitoring the response to step perturbations. The perturbations were introduced in AO closed-loop by moving one of AO stage by about 0.1\arcsec or 10 pixels on the NIX detector.
Two examples of those tests are given in Fig. \ref{fig:step_response}. The measurements are compared with a toy model of the QACITS closed-loop. On top of the two control parameters, the toy model is defined by the integration time and the time delay between measurement and correction. The first is given by the QACITS repetition rate, \ie 5 seconds. The second is not precisely known nor deterministic and was arbitrarily set in the toy model to 4 seconds to best match the measurements. This  time delay is the combined contributions of computational delays, communication delays between the QACITS and the asynchronous AO auxiliary loop, and the response time of the hardware itself.
Based on those tests, we select a proportional gain $K_p=0.25$ and an integrator gain $K_i=0.5$ which seems a good compromise between response speed and stability. With those parameters, the 0dB closed-loop error cut-off frequency is estimated to $\sim$0.02Hz, meaning that minute-scale pointing drift are efficiently rejected.

As a comparison, the VLT/NEAR and Keck/NIRC2 had respectively $\sim$30s  and $\sim$50s repetition rate. Therefore, the VLT/ERIS implementation provides a 5- to  10-fold improvement in terms of rejection bandwidth.

\subsection{Implementation in ERIS}
The QACITS algorithm has been implemented in Python, and is fully integrated to the ESO ERIS templates. A lot of the work has been dedicated to enable effort-less and robust operations.
The FPC operations are divided in two specific templates, one for acquisition and one for observation.

\begin{figure}[t]
    \centering
    \includegraphics[width=0.6\textwidth]{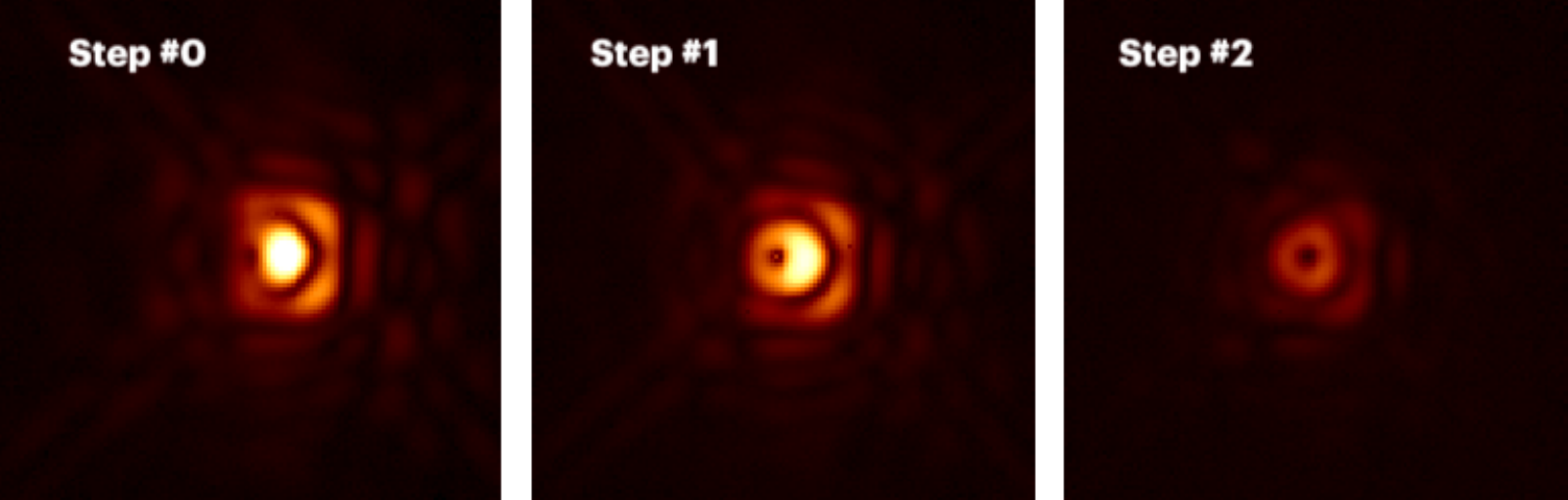}
    \caption{Improvement of the rejection of the on-axis star as the centering is improved at the end of the QACITS acquisition sequence. The images are sky subtracted and displayed with a square-root stretch. The left image has an offset of about 1 $\lambda/D$ between the star and the vortex, while the right image shows a well-centered star. The rejection of the stellar light is about a factor 2 initially (left) and reaches more than a factor 100 at the end of the sequence (right). The images are 1.7\arcsec across; for reference the FWHM in L band is 97 mas.}
    \label{fig:fpc_opt}
\end{figure}

After the regular telescope and AO target acquisitions,  the FPC acquisition template starts and executes the following steps:

\begin{enumerate}
    \item The target is first roughly centered onto the vortex mask (within $\sim 1 \lambda /D $).
    \item The target is offset outside the field-of-view and a set of dark and sky frames are acquired.
    \item Using a sky and dark frames with preset integration time, the vortex position is measured by fitting the vortex center glow, see also \sref{sec:vcg}.
    \item The target is moved in the field but far off-axis from the vortex center. An unsaturated image of the star is acquired to provide a photometric reference for  QACITS, and the actual offset of the star is measured by fitting a Gaussian profile.
    \item Based on the previous measurements, the star is moved onto the vortex. At the end of this step, the star is centered onto the vortex mask within approximately a 0.1 to 1 $ \lambda/D$ accuracy (\ie a few pixels).
    \item  The centering is then improved by an optimization sequence consisting of a few iterations of the QACITS loop, typically using shorter integration than the scientific acquisitions. The optimization sequence performs three iterations before moving to the observing template. The centering improvement is illustrated in \fref{fig:fpc_opt}.
\end{enumerate}
The acquisition time takes approximately 3.5 minutes.
The acquisition can be performed with either the regular Lyot stop or a Lyot stop coupled with a ND filter (extinction $> 1:60$) to avoid any saturation of the off-axis PSF on very bright stars.

\begin{figure}[t]
    \centering
        \includegraphics[width=0.7\textwidth]{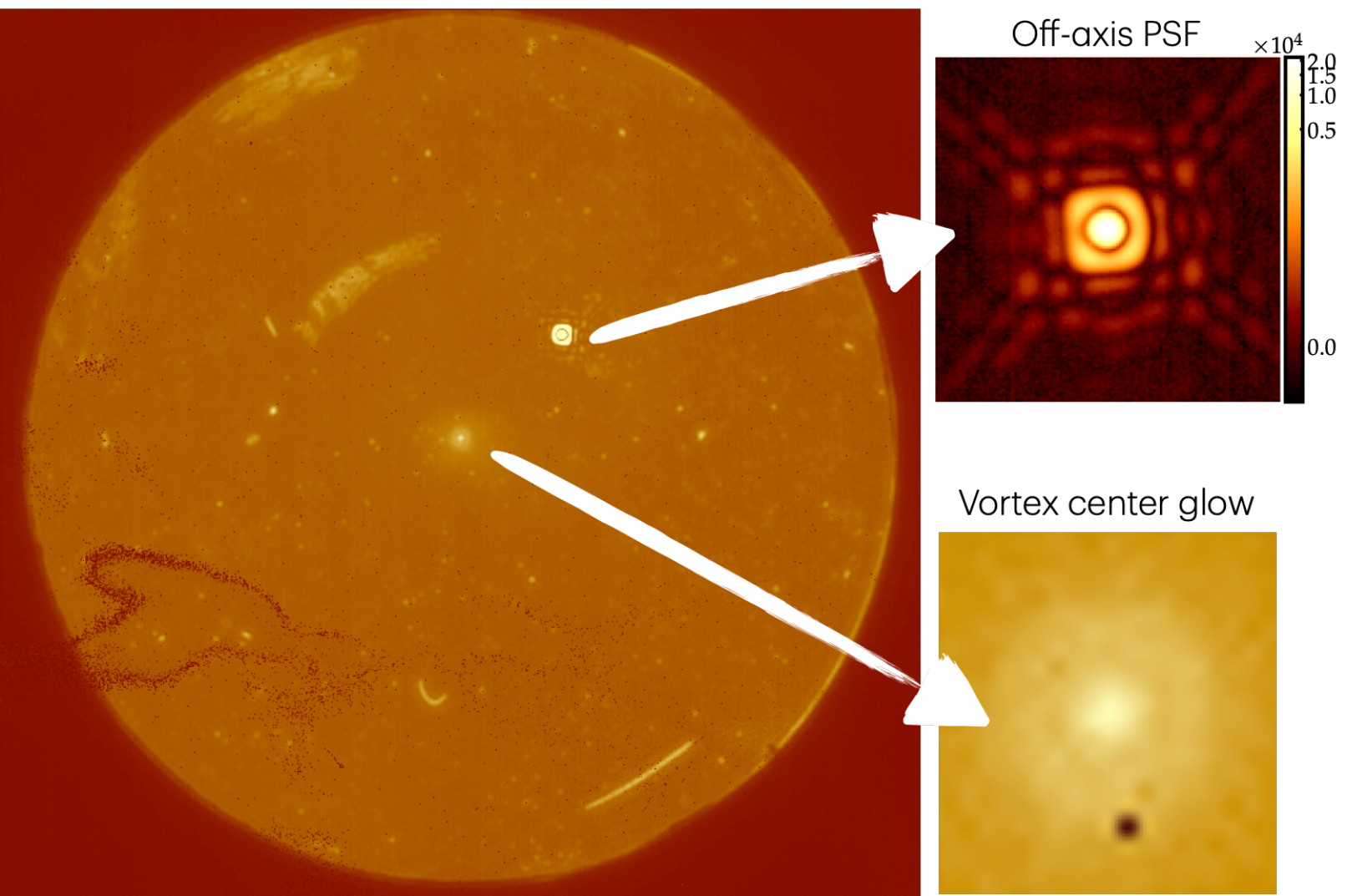}
    \caption{NIX sub-images during FPC acquisition. The large circle (left image) represents  the footprint of the vortex phase masks, equivalent to $\sim$16\arcsec diameter on-sky. The image features also cosmetics from the NIX detector, and dust contamination of the vortex phase masks. The image and the zoom insets display the diffraction-limited off-axis PSF and the vortex center glow, see also \sref{sec:vcg}.}
    \label{fig:fpc_vcg}
\end{figure}

Once the acquisition template is completed, the FPC science observing template starts. QACITS is enabled by default and measures pointing drift every 5 seconds based on the available data. One particularity is the monitoring of the vortex center discussed in the next section: anytime the observer requests a sky background, another background with a predefined integration time is also acquired, from which the vortex position is computed using the vortex center glow. This allows to keep track of any drift of the vortex position onto the NIX camera.

During the observing template, a simple live display shows the QACITS measurements. The QACITS measurements are  also logged in the generic instrument workstation logs. In the future, we plan to add quality metrics in the data headers, and possibly provide a fits table with all the QACITS measurements.

\subsection{Vortex position measurement}\label{sec:vcg}
Although the vortex mask is fully transparent, it typically produces a diffuse emission on top of the thermal sky background. This emission, observed with other instruments in the L, M, and N bands, is the result of thermal emission from the warm telescope structure and the environment around the entrance pupil. The vortex phase mask partially diffracts this thermal emission into the pupil image, creating a diffuse bright spot in the downstream image plane\cite{Absil+2016, Shinde+22}. If the AGPM was placed behind a cold pupil stop, we expect that this emission would disappear and would be replaced by a dark spot. Nevertheless, while this vortex center glow slightly increases photon noise locally, it allows the position of the vortex to be easily identified and monitored.

\begin{figure}[h!]
    \centering
    \includegraphics[width=0.8\linewidth]{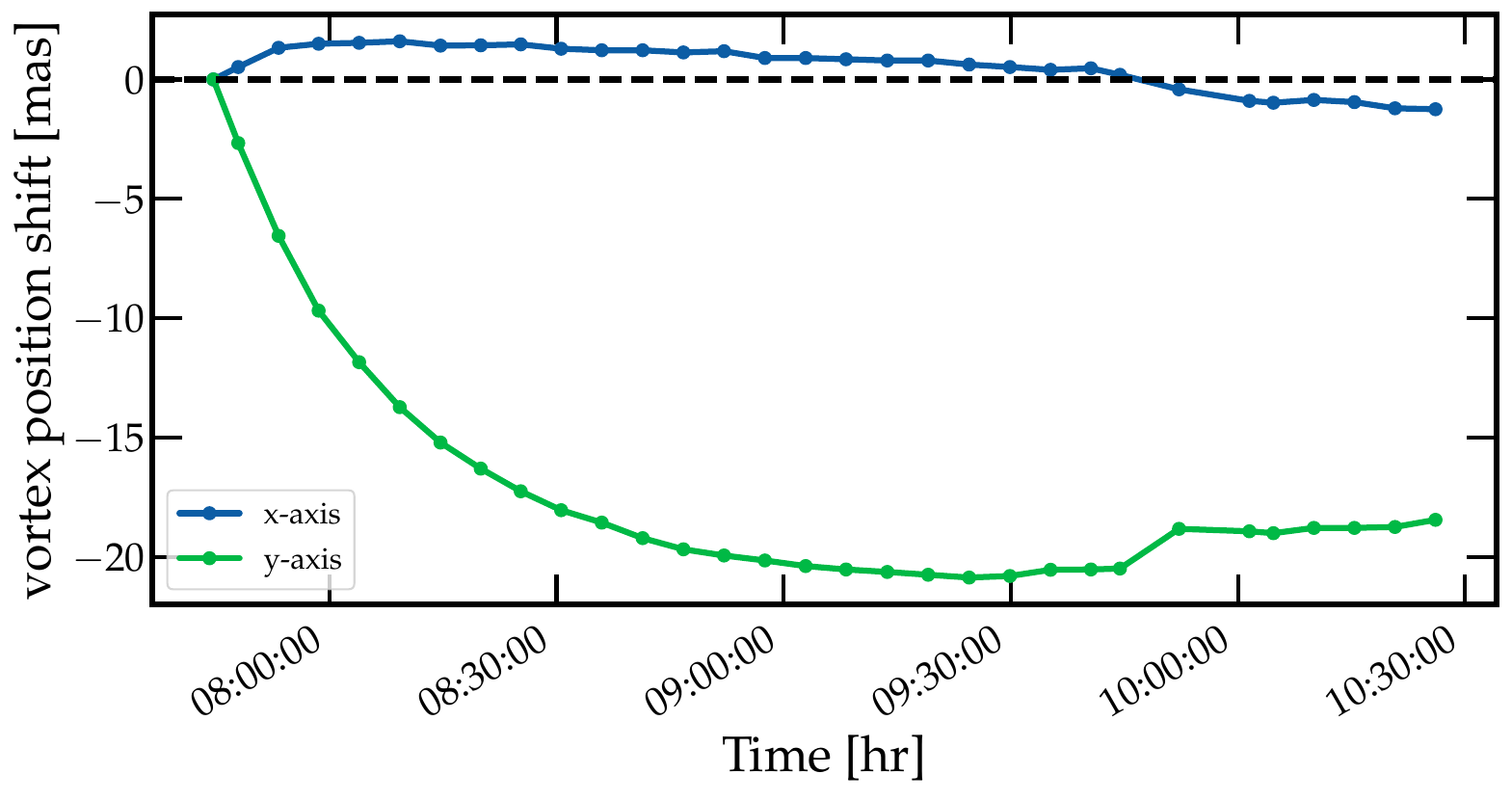}
    \caption{Representative vortex position drift in $x$ and $y$, as a function of time.  }
    \label{fig:vortex_drift}
\end{figure}

During commissioning, by monitoring off-line the vortex center glow, we noticed a non negligible drift of the vortex position with respect to the NIX detector, which is bound to directly impact the QACITS centering accuracy. To mitigate any drift on pointing accuracy, we implemented an automatic vortex center measurement on every sky frames, updating the target zero position of QACITS.
In \fref{fig:vortex_drift}, we illustrate the measured vortex position drift over a $\sim$2-hr observing sequence. We can notice a total drift of about 20~mas, or 0.2 - 0.3 $\lambda/D$.
Based on the fastest drift rates measured on-sky, we need to re-measure the vortex position approximately every 5~min.

\section{ON-SKY RESULTS}

\subsection{Pointing control performance}

\begin{figure}[t]
    \centering
    \includegraphics[width=0.3\linewidth]{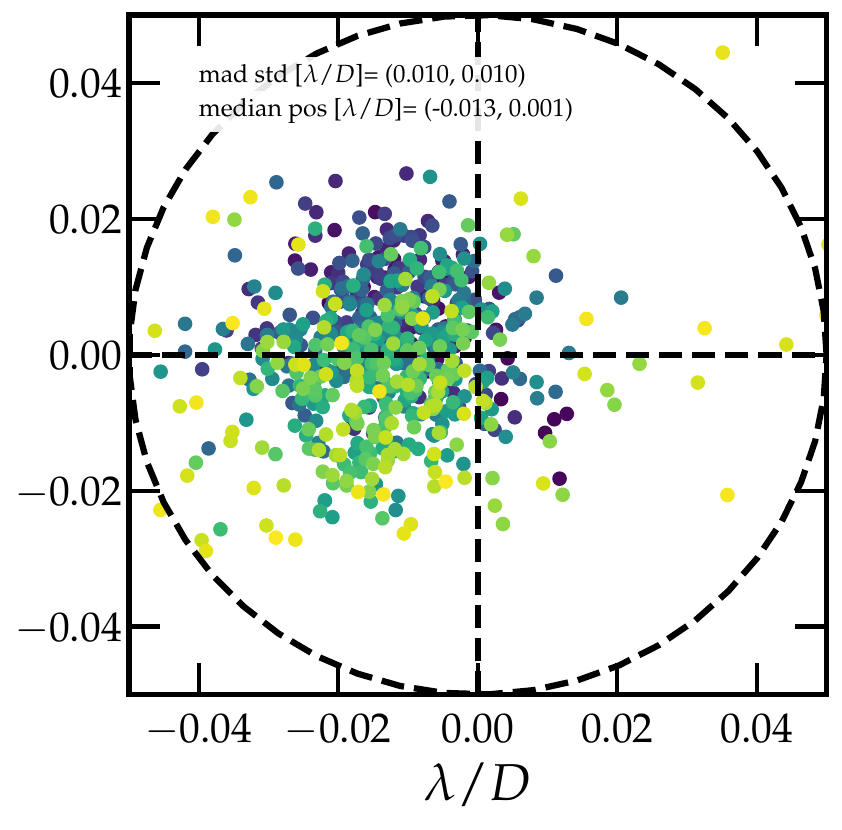}
    \includegraphics[width=0.3\linewidth]{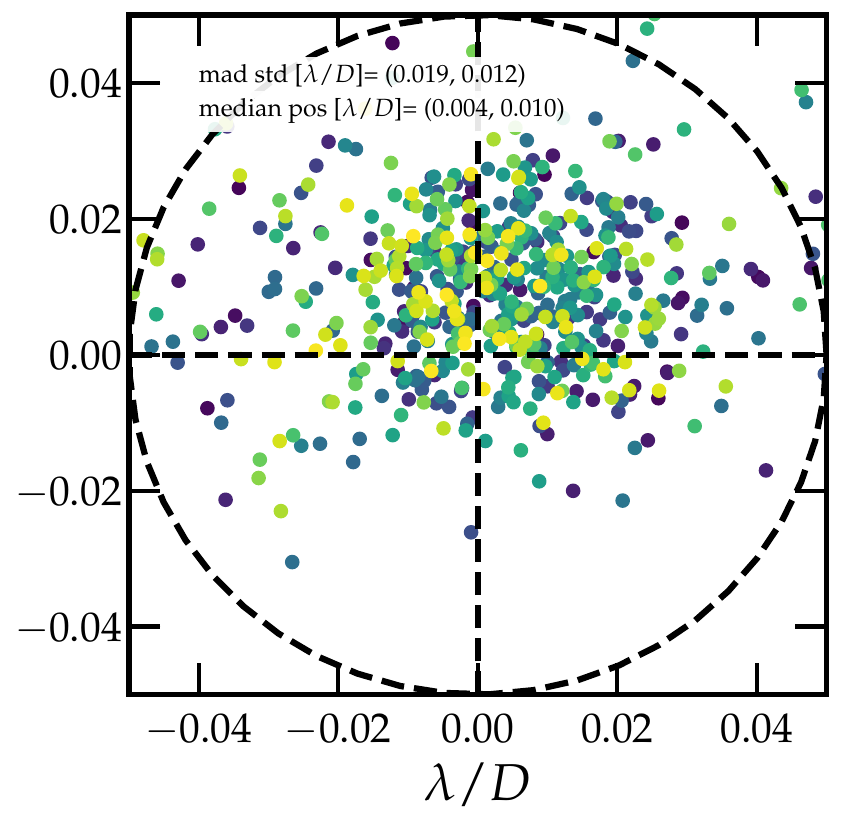}
    \includegraphics[width=0.3\linewidth]{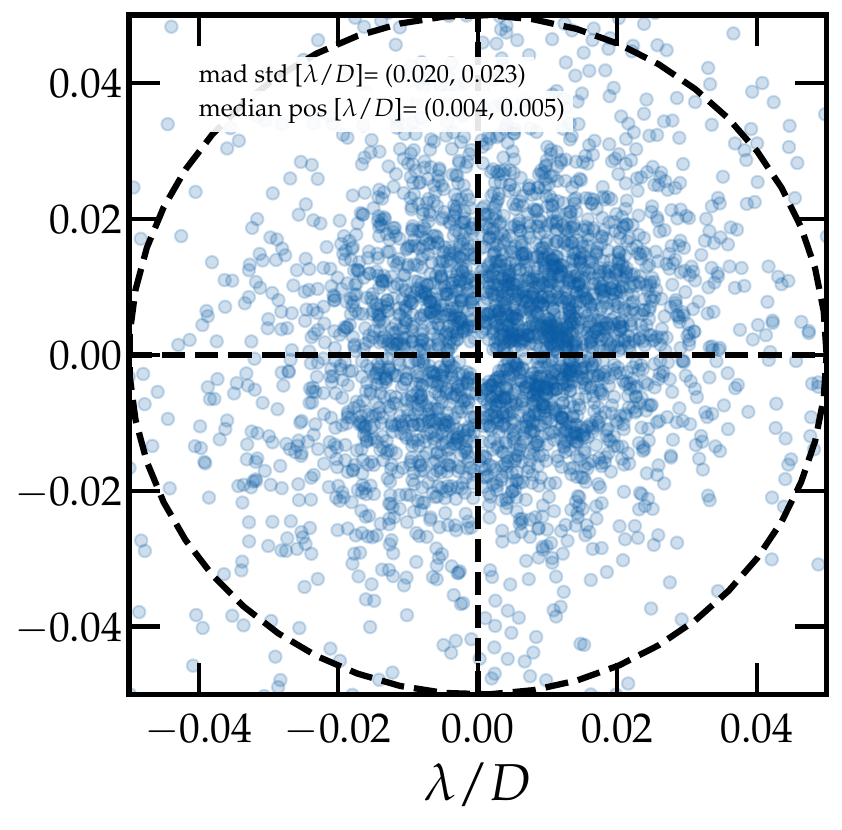}
    \caption{Residual pointing errors measured by QACITS. ({\it Left}) on i~Vel, a 'bright' target of $L-$mag=4.1, observed during approx. 60-min. ({\it Middle}) on GQ~Lup, a 'faint' target of $L-$mag=6.1, observed during approx. 70-min. ({\it Right}) On multiple targets during an entire night of FPC operation. }
    \label{fig:pointing}
\end{figure}

We illustrate in \fref{fig:pointing} the QACITS pointing performance obtained on-sky based on two $\sim$1hr observations performed during commissioning in May 2023, and an entire night during Period 113 (Spring 2024).
The first target is i~Vel with a mag$_L$=4.1 (`bright case') observed during 62-min where the pointing precision (robust standard deviation) is $\sim$0.01$\lambda/D$.
The second target is GQ~Lup with a  mag$_L$=6.1 (`faint case') observed during 70-min where the pointing accuracy is $\sim 0.02 \lambda/D$.
Finally, we  illustrate an entire night of focal plane coronagraphic observations of multiple targets where the overall pointing precision is also $\sim 0.02\lambda/D$, testifying the good robustness of QACITS.

\begin{figure}[t]
    \centering
    \includegraphics[width=0.3\linewidth]{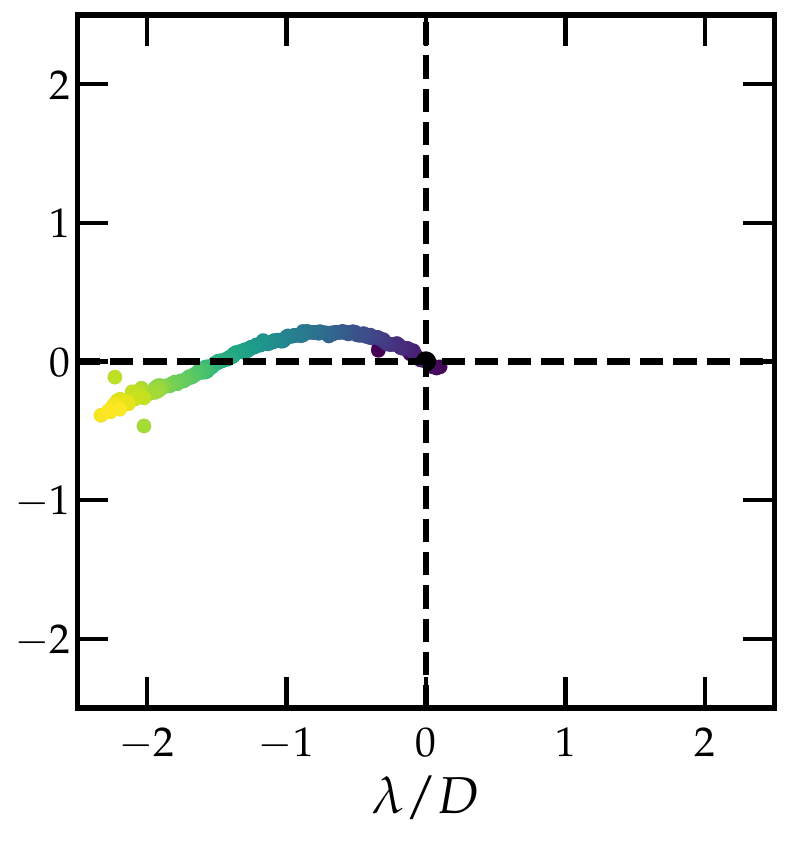}
    \caption{Time evolution (color coded) of the QACITS total command measured on i~Vel over 1-hr. The curve is typical of differential atmospheric refraction, with a meridian passage after 15-20 min. }
    \label{fig:drift}
\end{figure}

Systematic shift, such as the one observed on i~Vel of about 0.01$\lambda/D$, can be explained by the star pointing drift. In \fref{fig:drift}, we plot the total command of QACITS over the 1hr sequence. We can observe a drift of about 20 pixels/1hr or 4mas/min. Considering the closed-loop control and its inherent delays, this is consistent with an observed shift of about 0.01 $\lambda/D$. It goes without saying that without QACITS, this drift would strongly degrade the vortex performance.

Based on the different results obtained on-sky, we can conjecture that QACITS would still work up to mag$_L = 8$ (although at a reduced pointing accuracy), and the control loop would eventually break between magnitude 8 and 9. We should stress however that the usefulness of coronagraphic observations for objects fainter than magnitude $\sim$ 9 is questionable. Indeed, in this regime, we expect the contrast to be limited by background noise at most separations, making normal imaging the preferred mode.

\subsection{Coronagraphic performance}
\subsubsection{Raw contrast}
While the vortex coronagraph is expected to perfectly reject the starlight (with a circular pupil and flat wavefront), four main effects degrade its rejection in practice \cite{Serabyn+17, Absil+2016} : (i) the diffraction of the central obscuration and the spiders, (ii) the pointing errors, (iii) the adaptive optics residuals and other sources of residual wavefront errors (e.g., NCPA), (iv) the intrinsic imperfections of the vortex phase mask.
In the following, we provide typical order of magnitude of the integrated leakage due to those effects:

\begin{enumerate}
    \item The fractional leakage due to the central obscuration can be approximated to first order\cite{Serabyn+17} by $(d/D)^2$. With the VLT central obscuration $d$ and telescope diameter $D$, the estimated leakage is $\sim0.05$. However, this approximation does not take the spiders into account nor the Lyot stop blocking a large fraction of this leakage. To have a more realistic estimate, we perform a Fourier-based simulation of the ERIS vortex coronagraph and compute a leakage of $\sim0.016$.
    \item Next, small pointing errors $\theta$ lead to a leakage given\cite{Huby+15} by $(\pi \theta / (\lambda/D))^2 / 8$. Based on the pointing precision obtained with QACITS ($\sim0.01-0.02\lambda/D$), the leakage would be $<5e-4$. However, this only represents the slow-evolving pointing drifts and we may expect jitter residuals from atmospheric turbulence and telescope vibration. If we reasonably assume a fast residual jitter of 10mas, the leakage term would be $\sim 0.012$.
    \item Wavefront errors will also produce leakage within the geometric pupil. For a good AO correction, this can be approximated by $1-S$, with $S$ the Strehl ratio. With ERIS, we have about 95\% Strehl in L band, thus the associated leakage is approximately 0.05.
    \item The intrinsic imperfection of the vortex phase mask translates essentially in a chromatic leakage, i.e. a fraction of the light is not affected by the vortex phase ramp and thus `leak' inside the geometric pupil. The ERIS components have been tested on our VODCA bench at ULiège, and perform very well with rejection ratios above $1000:1$, see Tab. \ref{tab:agpm}. Thus the leakage associated to the mask themselves is negligible and $<0.001$.
\end{enumerate}

\begin{figure}[ht!]
    \centering
    \includegraphics[width=0.7\linewidth]{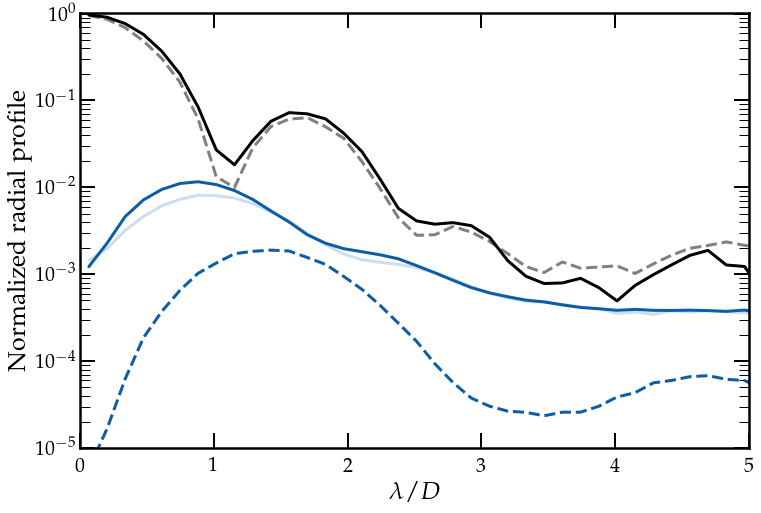}
    \caption{Radial intensity profiles with the star centered on the vortex phase mask (blue) or off-axis (black). Simulated profiles (dashed lines) are also computed for comparison, assuming perfect wavefront. While the solid blue line represents the average profile, the light blue one corresponds to the best profile (best rejection) obtained during the i~Vel observation.}
    \label{fig:radial-profile}
\end{figure}

Summing up the various effects, we can expect a total integrated leakage of about 0.06 - 0.08.
In \fref{fig:radial-profile}, we show the  radial profiles obtained on i~Vel in May 2023, \ie the off-axis PSF and the on-axis coronagraphic PSF and a comparison to Fourier-based simulations (under perfect conditions, \ie using the actual ERIS pupil and Lyot stops but without any wavefront or pointing errors). The measured integrated leakage is here $\sim 0.065$ indicating the good match with our rough error budget (and the good Strehl and low residual jitter).

Beyond the absolute value of the integrated leakage, the stability of the leakage and the post-coronagraphic PSF is key to ensure an optimum PSF subtraction in post-processing. Assuming that the Strehl ratio is stable within 0.5\%, we can derive a pointing stability requirement on par with the AO residual leakage. The derived stability requirement is $\sim 0.05 \lambda/D$ at L band on the VLT telescope, and can be considered as a reasonable precision requirement for  QACITS.

Other effects such as misalignment (e.g. Lyot stop) or NCPA can also significantly degrade the performance, if not cautiously tracked down. In the case of ERIS, as far as we can judge, these effects do not play a significant role on the raw contrast, but may still significantly affect the sensitivity limits after Angular Differential Imaging (ADI) processing. Finally, this integrated leakage error budget does not address where the different leakage terms end up in the focal plane (e.g. low order distortion of the diffraction pattern, or speckles) nor their interference effects (which may lead to semi-static "pinned" speckles).

\subsubsection{Post-processed contrast in \textit{L'}}
During commissioning, the FPC on-sky contrast curves have been derived in L band on two different objects: i~Vel (mag$_L$ = 4.1) and GQ~Lup (mag$_L$ = 6.1). The details of the observations and the results are presented below.

\begin{figure}[t]
    \centering
    \includegraphics[width=0.75\linewidth]{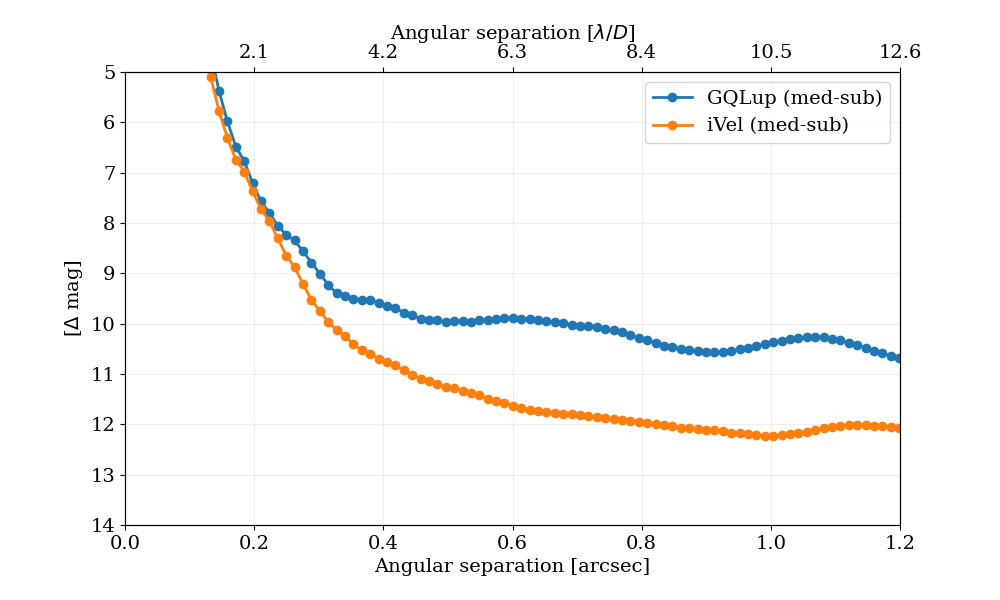}
    \caption{5-sigma post-processed contrast curves obtained with the vortex focal plane coronagraph and median PSF subtraction. ({\it Blue}) GQ~Lup
with a total ADI sequence of 70-min and a parallactic angle rotation of 59 degrees. (Orange) i~Vel with a total
ADI sequence of 62-min and a parallactic angle rotation of 39 degrees. With annular PCA, the detection limit goes down by another magnitude, reaching down to $\Delta\sim13 {\rm mag}$ for i~Vel.}
    \label{fig:cc}
\end{figure}

\begin{figure}[t]
    \centering
    \includegraphics[width=0.32\linewidth]{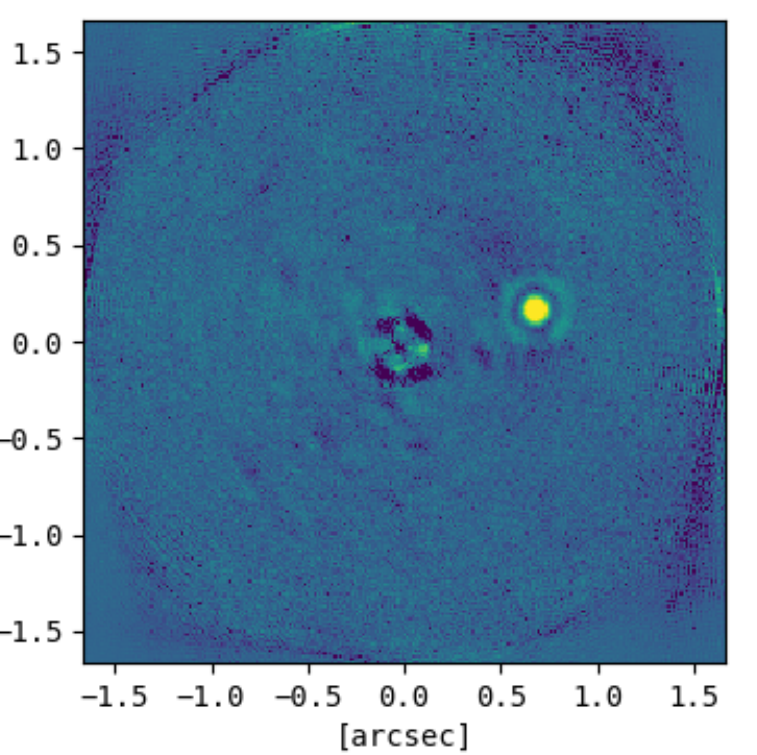}
    \caption{Post-processed image of GQ~Lup using median subtraction, revealing the companion, GQ~Lup~b, at about 0.6\arcsec. }
    \label{fig:gqlup}
\end{figure}

\textit{GQ~Lup} was observed in a Object-Sky-Object-... sequence with 400 DIT of 0.3s on Object and 10 DIT of 0.3s on Sky, providing a net exposure time of 52-min on target (after frame selection). The total exposure time, including sky frame, was 1h1m and the total execution time was 1h23m. The total ADI sequence amounts to 58.6 degrees of parallactic angle rotation. The $R$-mag (AO guide star) of GQ~Lup is 10.3.
The data reduction used the \texttt{VIP} package\cite{Christiaens+23} and includes dark subtraction, flat correction, bad pixel removal, recentering of frames, PCA sky subtraction, absolute recentering, and median-subtraction or annular PCA for post-processing. The resulting contrast curve is given in Figure \ref{fig:cc}. This curve represents the 5-sigma contrast, corrected for small sample statistics\cite{Mawet+14}. Prior to computing the contrast curve, the companion GQ~Lup~b is removed.
The  curve accounts for the off-axis transmission of the vortex, as illustrated in Figure \ref{fig:eris-simu}.
We also produce an ADI post-processed image, see \fref{fig:gqlup}, providing an exquisite detection of the GQ~Lup companion at about 0.6\arcsec.

\textit{i~Vel} was observed following the same strategy as GQ~Lup, with a net exposure time of 60-min on target.  The data processing and post-processing follows the same principle. The resulting contrast curves are given in Figure \ref{fig:cc} .
The performance obtained are similar to GQ~Lup up to about 2.5 $\lambda$/D ($\sim0.25\arcsec$). At larger separation, the contrast is better with i~Vel.
A brighter star means improved AO performance, increased pointing accuracy with QACITS ($\sim$0.01 $\lambda$/D with i~Vel instead of $\sim$0.02 $\lambda$/D for GQ~Lup), and a deeper background-limited sensitivity in terms of contrast, which account for the difference in the contrast curves beyond 2.5 $\lambda$/D.

\subsubsection{Comparison with NACO}
In an attempt to compare apples with apples, we selected four datasets in the ESO archive taken with the vortex coronagraph of NACO. The datasets are chosen to have similar target magnitudes and observing conditions.
In addition to the i~Vel ERIS observations with a 57 minutes integration time and 39 degrees field rotation, we have HD~11171 (2 hours 2 minutes; 108.7 degrees), HD~188228 (1 hours 46 minutes; 63 degrees), HD~71155 (47 minutes, 48 degrees), HD~98058 (1 hours 44 minutes, 88deg).
All NACO and ERIS observations were reduced with the same post-processing pipeline using \texttt{PynPoint}\cite{Stolker+2019}. The pipeline includes basic calibrations (dark subtraction, flat correction, bad pixel removal, recentering of frames) as well as a PCA-based\cite{Amara+2012,Soummer+2012} PSF subtraction. The number of PCA components is optimized to reach the deepest contrast for each separation.
Afterwards we compute contrast curves based on a t-test \cite{Mawet+14} using \texttt{applefy}\cite{Bonse+2023}.

\begin{figure}[h!]
    \centering
    \includegraphics[width=0.75\linewidth]{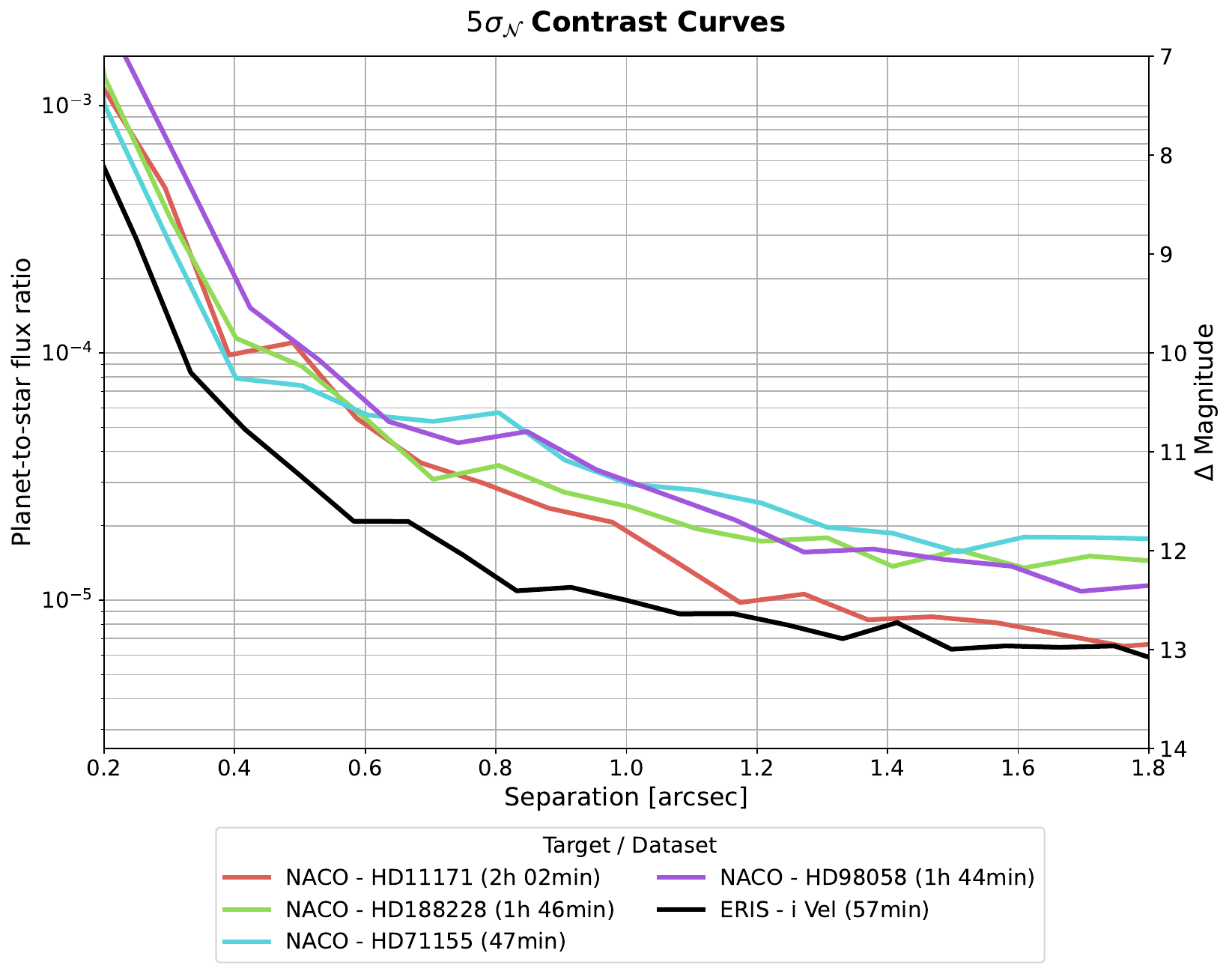}
    \caption{5-sigma contrast curves, comparing a set of representative NACO observations with the observation of i~Vel with ERIS. The NACO observations have different integration times and field rotations, but overall, the ERIS contrast curve is about 1 mag better than NACO.}
    \label{fig:cc_naco}
\end{figure}

The results are given in \fref{fig:cc_naco}.
Overall, considering the variety of integration times and field rotations (which globally favor the NACO observations), the results indicate a gain of approximately 1 magnitude for ERIS at all separations, both in the speckle-limited and background-limited regimes.
This comparison provides a first indication of the excellent performance of the ERIS AO system and the FPC coronagraph, and is promising for forthcoming science observations.

\section{CONCLUSIONS}

In this contribution, we have presented the newly commissioned focal plane coronagraphic mode of ERIS, featuring vortex phase masks. We have reviewed its design,  pointing control with QACITS,  template implementation, and  on-sky results in terms of pointing stability, raw coronagraphic performance and post-processed detection limits.

While the QACITS algorithm is straightforward in theory, its implementation can be challenging due to its interface between high-contrast imaging, instrument control, and adaptive optics. This typically requires the intervention of multiple experts.
Access to a mid-infrared calibration source can facilitate the algorithm's integration and enable early testing, e.g. during system AIV, significantly reducing the required on-sky time at commissioning. This early integration approach is  foreseen for the ELT/METIS instrument.

The implementation of the ERIS FPC mode is now mostly finalized, enabling robust vortex operations. Its full integration into the ESO environment  provides an easy access for the scientific community.
A good pointing stability is now systematically achieved within 0.02 $\lambda/D$ or better, and with a rejection bandwidth of about 0.02Hz. In an hour-long observation, we obtain a 5$\sigma$ background level of $L^{\prime} > 17$. However, how close to the star this performance level can be reached  strongly depends  on the residual pointing and wavefront errors. In both speckle and background-limited regimes, the first ERIS results show a promising improvement of about 1-mag over previous NACO observations.
While, in this proceeding, we have primarily focused on the L band vortex, the M band mode has just been commissioned and we have demonstrated similar pointing stability with QACITS. Both vortex modes are thus ready for scientific exploitation.

\acknowledgments 
GOX thanks Valentin Christiaens for discussion on the ERIS data reduction using the \texttt{VIP} Python package.
GOX and OAb acknowledges funding from the European Research Council (ERC) under the European Union’s Horizon 2020 research and innovation programme (grant agreement no. 819155), and from the Wallonia-Brussels Federation (grant for Concerted Research Actions). This work has been carried out within the framework of the NCCR PlanetS supported by the Swiss National Science Foundation under grant 51NF40\_205606. JH acknowledges the financial support from the Swiss National Science Foundation under project grant number 200020\_200399.

\bibliography{report} 
\bibliographystyle{spiebib} 

\end{document}